\documentclass[10pt,twocolumn]{IEEEtran} 

\usepackage{cite}
\usepackage{graphicx}
\usepackage{mathtools}
\usepackage{amsmath,amssymb,amsfonts}
\usepackage{svg}
\usepackage{amssymb}

\usepackage[absolute]{textpos}

\usepackage{lipsum}
\usepackage{algorithm}
\usepackage{algpseudocode}
\usepackage{graphicx}
\usepackage{booktabs,amsfonts,dcolumn}
\usepackage[T1]{fontenc}
\usepackage{pifont}
\usepackage{textcomp}
\usepackage{hyperref}
\usepackage{enumitem}

\def\BibTeX{{\rm B\kern-.05em{\sc i\kern-.025em b}\kern-.08em
    T\kern-.1667em\lower.7ex\hbox{E}\kern-.125emX}}

\usepackage{tikz}
\usepackage{subcaption}
\usepackage{color}
\usepackage{siunitx}
\usepackage{chngcntr}
\usepackage{bm}
\usepackage{ifthen}
\usepackage{soul}
\usepackage{tabularx}
\usepackage[acronym]{glossaries}

\newcommand{\comment}[1]{}
\newboolean{paper}
\setboolean{paper}{true}

\newacronym{ann}{ANN}{artificial neural network}
\newacronym{mac}{MAC-OP}{multiply and accumulate operation}
\newacronym{td}{TD}{time-domain}
\newacronym{tdmac}{TD-MAC}{time-domain MAC}
\newacronym{snr}{SNR}{signal-to-noise ratio}
\newacronym{adc}{ADC}{analog-to-digital converter}
\newacronym{tdc}{TDC}{time-to-digital converter}
\newacronym{vmm}{VMM}{vector-matrix-multiplication}
\newacronym{cim}{CIM}{compute-in-memory}
\newacronym{cnn}{CNN}{convolutional neural network}
\newacronym{lsq}{LSQ}{learned step size quantization}
\newacronym{enob}{ENOB}{effective number of bits}
\newacronym{evpv}{EVPV}{expected value of the process variance}
\newacronym{vhm}{VHM}{variance of the hypothetical means}
\newacronym{csnr}{$\eta_\text{ESNR}$}{SNR adjusted energy efficiency}
\newacronym{pnr}{P\&R}{place and route}
\newacronym{inl}{INL}{integral nonlinearity}    
\DeclareMathOperator{\EX}{\mathbb{E}}

\begin{document}
\title{\vspace{0.8cm}Merits of Time-Domain Computing for VMM - A Quantitative Comparison} 

\author{\large Florian Freye,  Jie Lou, Christian Lanius and Tobias Gemmeke \\ 
 Chair of Integrated Digital Systems and Circuit Design, RWTH Aachen University, Germany\\
 E-mail: freye@ids.rwth-aachen.de\thanks{This work has been funded in part by the Federal Ministry of Education and Research (BMBF) through the NEUROTEC Project under Project 16ME0399.}}


\maketitle
\thispagestyle{empty}\pagestyle{empty}
\begin{textblock*}{\textwidth}(1cm,0.5cm) 
© 2024 IEEE. Personal use of this material is permitted. Permission from IEEE must be
obtained for all other uses, in any current or future media, including
reprinting/republishing this material for advertising or promotional purposes, creating new collective works, for resale or redistribution to servers or lists, or reuse of any copyrighted component of this work in other works. This paper was  accepted at the 25th International Symposium on Quality Electronic Design (ISQED) 2024. DOI: \href{https://doi.org/10.1109/ISQED60706.2024.10528682}{10.1109/ISQED60706.2024.10528682} \vspace{0.1cm}

\end{textblock*}

\begin{abstract}
\Gls{vmm} accelerators have gained a lot of traction, especially due to the rise of \glspl{cnn} and the desire to compute them on the edge.
Besides the classical digital approach, analog computing has gone through a renaissance to push energy efficiency further.
A more recent approach is called \gls{td} computing.
In contrast to analog computing, \gls{td} computing permits easy technology as well as voltage scaling.
As it has received limited research attention, it is not yet clear which scenarios are most suitable to be computed in the \gls{td}.
In this work, we investigate these scenarios, focussing on energy efficiency considering approximative computations that preserve accuracy.
Both goals are addressed by a novel efficiency metric, which is used to find a baseline design.
We use SPICE simulation data which is fed into a python framework to evaluate how performance scales for \gls{vmm} computation.
We see that \gls{td} computing offers best energy efficiency for small to medium sized arrays.
With throughput and silicon footprint we investigate two additional metrics, giving a holistic comparison.
\end{abstract}
\section{\bf Introduction}
\begin{figure}[thb]
    \centering
    \includegraphics[width=0.85\columnwidth]{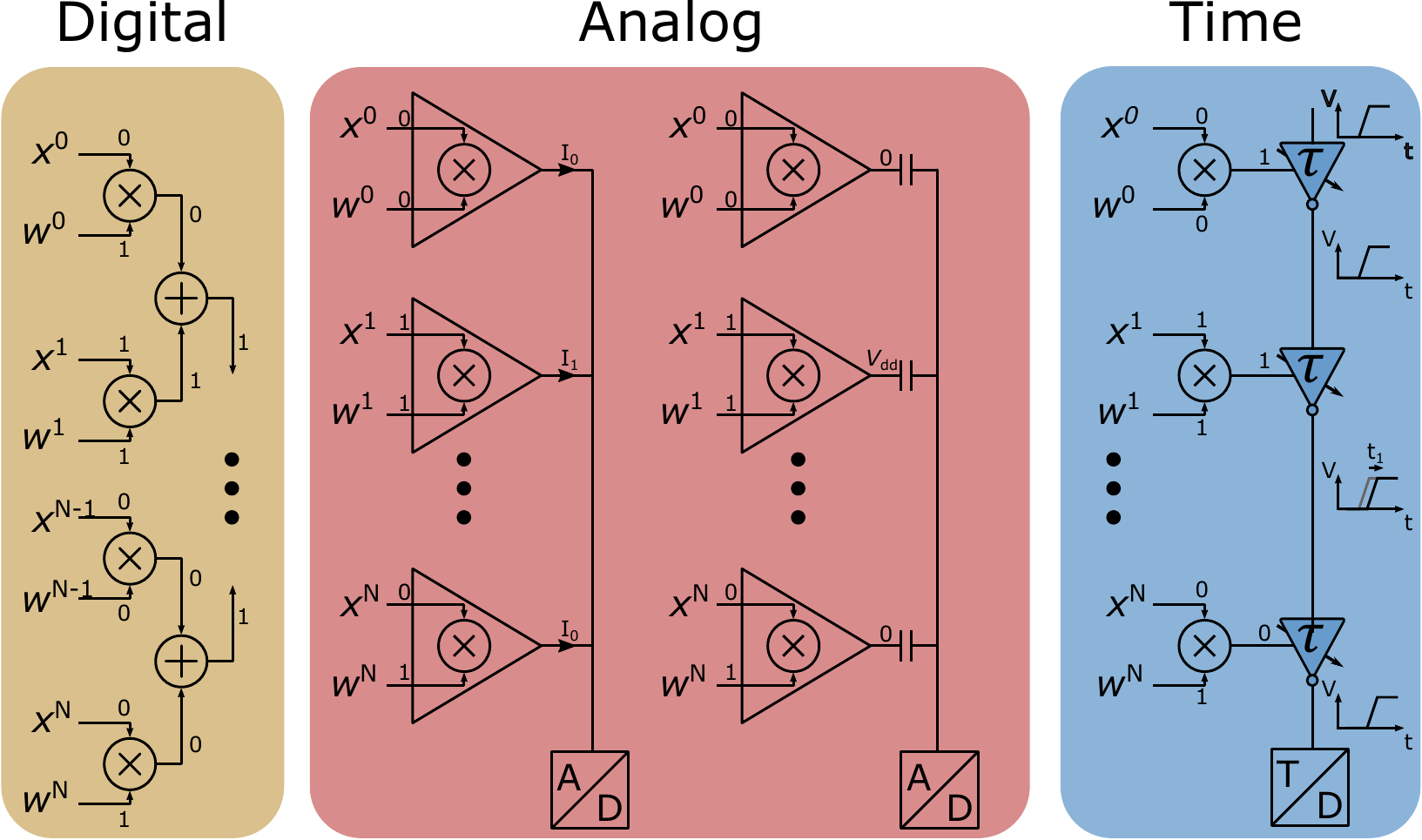}
    \caption{Overview over different compute domains.}
    \label{fig:domain_overview}
\end{figure}

Due to the fast adoption of \glspl{ann}, a pressing research interest lies in the decrease of energy consumption to allow the computation on the edge. Besides memory transfer, the \gls{vmm} cost is a pressing research area as its building block, the \gls{mac}, is performed multiple billions of times in a single inference.
While classical digital computations have relied on basic adder trees, advances in efficiency have been mostly driven by technology scaling  or custom implementations of the \gls{mac} (Fig.~\ref{fig:domain_overview} left) \cite{Chih.2021, Wang.2022}.

Analog computing schemes have entered the discussion, as they have shown huge improvements in energy and area efficiency.
The analog domain can be further split up into current domain and charge domain  (Fig.~\ref{fig:domain_overview} middle) \cite{Chen.2021, Sharma.2021, Sinangil.2021, kneip.2023}. 
In current domain computing, a capacitive load is discharged by multiple drivers in parallel. 
Here, the height and duration of the current is modulated as function of the given input in a classic pulse-width modulation scheme. 
The residual charge represents the result of the computation.
A second option lies in accumulating currents resulting from a fixed voltage being applied to a series of parallel connected resistors of adjustable magnitude.
An \gls{adc} is then used to convert the result of the \gls{vmm} to the digital domain.
Current domain computing offers extremely high efficiency and compute density at the expense of reduced \gls{snr}.
Especially for \gls{cim} implementations using SRAM or crossbar arrays, current varies greatly due to the inherent device variations, which are amplified by the typically reduced voltage on the access transistor \cite{kneip.2023}.

Instead of connecting all driver outputs to add their currents, charge domain computing uses capacitive coupling to combine all multiplication results. 
The injected charges accumulate on the output connection.
In contrast to current domain computing, a direct integration into standard SRAM or ReRAM crossbar arrays is not possible, thus increasing area. 
Low mismatch for on-chip capacitors significantly increases the \gls{snr} of this architecture, leading to rising popularity of charge domain computing.
Despite these benefits, analog computing introduces challenges, such as \gls{adc} energy consumption, re-design effort in technology migration and poor voltage scaling behavior.

A promising alternative that aims to address these limitations is the so-called \gls{td} computing, which utilizes the time of signal transitions to encode discrete values  (Fig.~\ref{fig:domain_overview} right).
Here, additions are achieved by delaying the transition in proportion to the value of the addend.
At the end of a \gls{vmm}, a conversion to the digital domain is achieved using a \gls{tdc}.
In contrast to analog computing schemes, \gls{td} computing can be composed using only standard cells, easing technology transfer and voltage scalability \cite{Lou.2022}.

\begin{figure}[hbt]
    \centering
    \includegraphics[width=0.85\columnwidth]{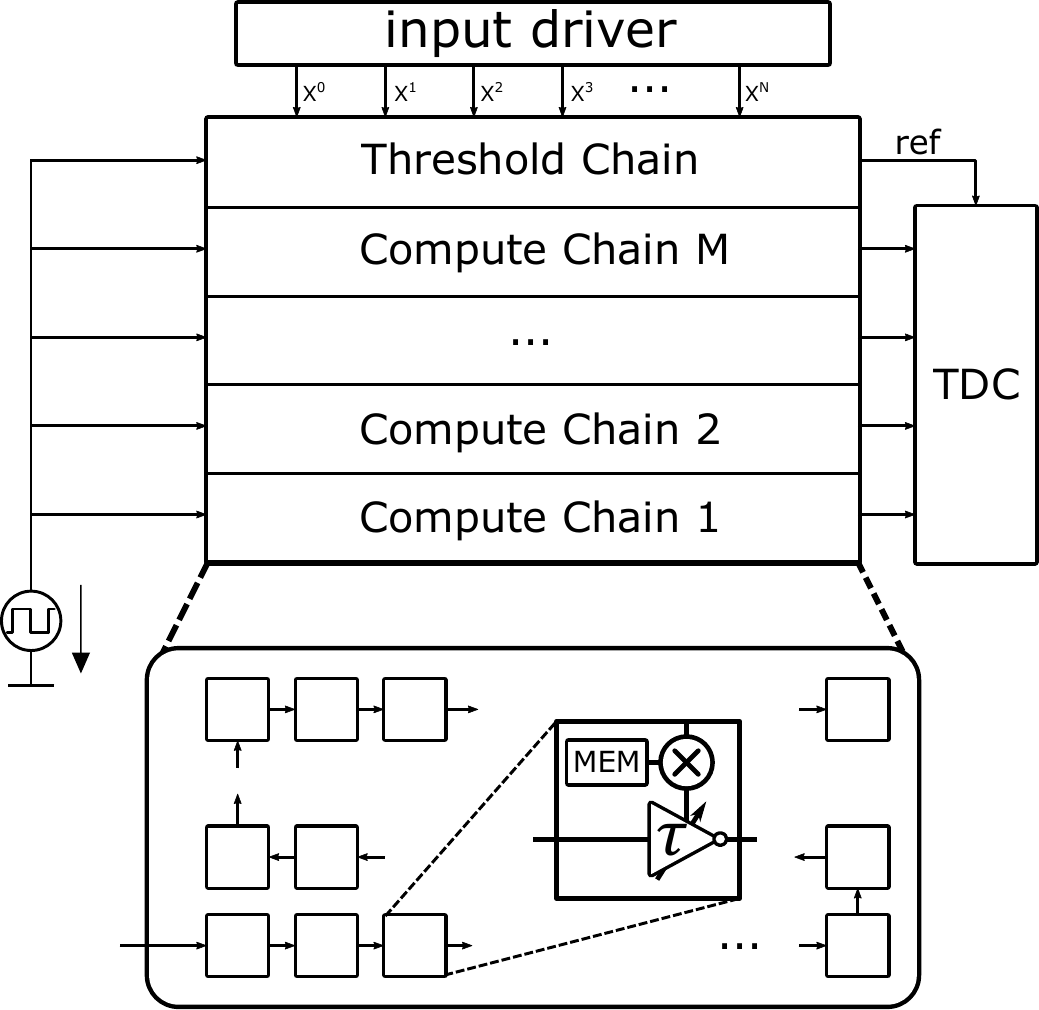}
    \caption{Weight static \gls{td} \gls{vmm} system overview.}
    \label{fig:td_overview}
\end{figure}

An overview of a typical \gls{td} macro for \gls{vmm} is presented in Fig. \ref{fig:td_overview}. 
Here, the matrix is static and saved in the MEM blocks, which is a typical \gls{cim} approach to \gls{vmm} and the vector values are entered as $x^i$ and shared between $M$ chains, to utilize the periphery more efficiently.
To implement a more favorable aspect ratio, the compute chains are arranged in a snake-like pattern (Fig. \ref{fig:td_overview} bottom).

While \gls{td} computing has proven to be able to provide good energy efficiency, compute accuracy is reduced by effects of device mismatch, random noise, nonlinearity and differences in wire lengths.
While the latter can be tackled by means of structured macros and custom routing \cite{Lanius.2023}, other aspects are often overlooked when reporting \gls{ann} classification accuracy only.
The authors of \cite{amaravati.2019} draw a comparison between \gls{td} and digital computation, which is performed based on \gls{pnr} results without consideration of the errors introduced by the computation.

In contrast, considering the literature on analog domain, comparisons accounting also for compute accuracy can be found which utilize relationships in SNR and energy consumption.
The author of \cite{Sar.1998} compares analog with digital computing to find the secret of the outstanding energy efficiency of the brain.
Here, analog computation is considered in time continuous fashion and utilizes transistors in sub-threshold operation.
In \cite{Murmann.2021}, energy efficiency of the more modern application of analog computing to \gls{vmm} is investigated for different bitwidths as well as array dimensions.
The analysis of the energy consumption of the logic cell and the capacitor are estimated using analytical models and the \gls{adc} energy consumption is estimated using an envelope of historic publications \cite{adc_survey}.
In \cite{houshmand.2023}, these methods are transferred from charge to current domain computing and an additional analytical model for digital computing is introduced. 

\subsection{\bf Paper Contributions}
In this work, we bring accuracy and energy efficiency together for the \gls{td}, assessing its efficiency as function of vector size in comparison to analog and digital computing. 
More specifically, this work makes the following contributions:
\begin{itemize}[leftmargin=20pt]
  \item A novel metric to optimize \gls{tdmac} cells is presented.
  \item A generic \gls{tdmac} cell is introduced, that allows the configuration from single to multi bit operations at tunable accuracy. To form a good baseline for the \gls{td}, we also provide a scalable \gls{tdc}.
  \item We quantitatively model \gls{vmm} computation in all three compute domains and investigate their strengths and weaknesses.
  \item Precise and approximate computations are considered.
\end{itemize}

\section{\bf TD-MAC cell} \label{sec:tdmac}
\begin{figure}[htb]
\centering
\tikz{
\node[anchor = south] at (0,0) (p1){\includegraphics[width=\columnwidth]{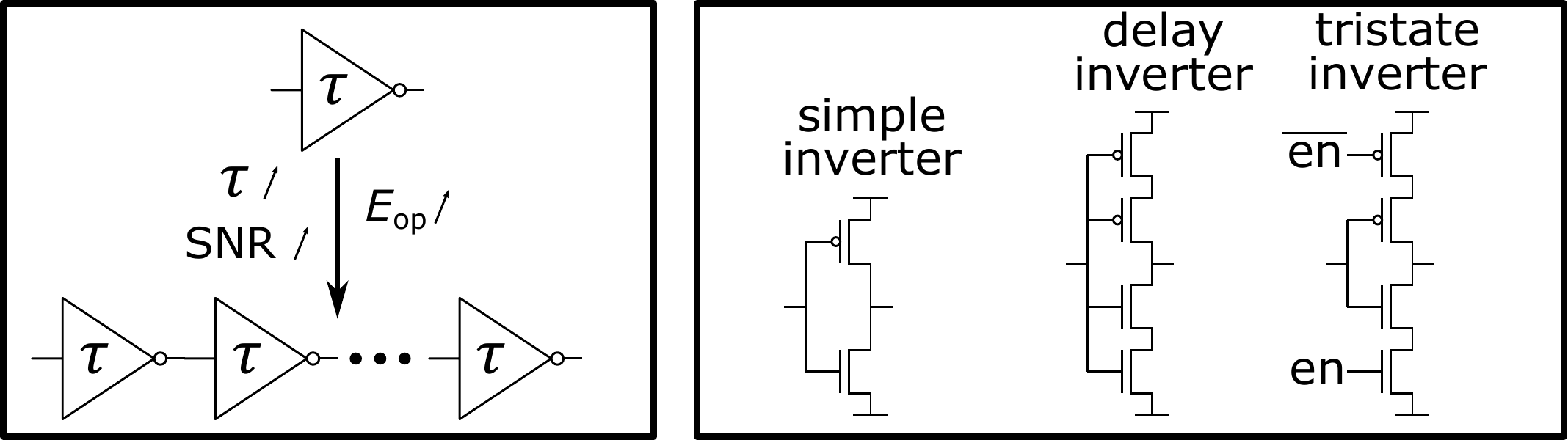}};
\node at (-4,2.4) {a)};
\node at (0,2.4) {b)};
\node at (-4,-0.5) {c)};
\node[anchor = north] at (0,0.2) {\includegraphics[width=0.79\columnwidth]{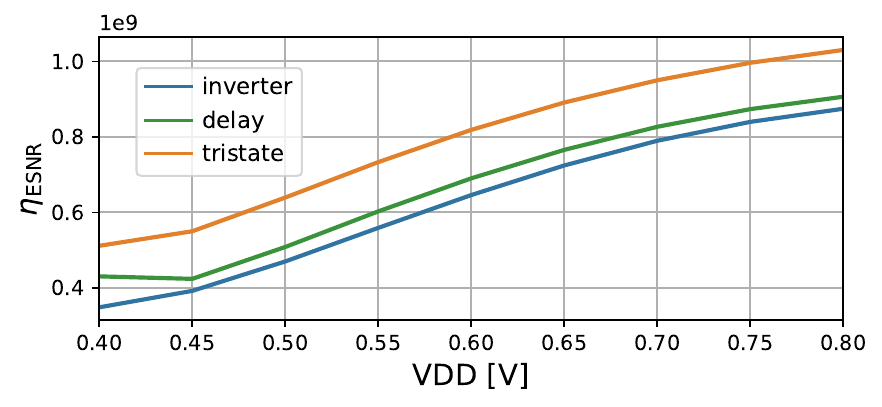}};
\node at (-4,-0.5) {c)};
}
\caption{(a) Cascading behavior. (b) Basic delay cells. (c) \gls{csnr} comparison of cells in b.}
\label{fig:cell_comparison}
\end{figure}
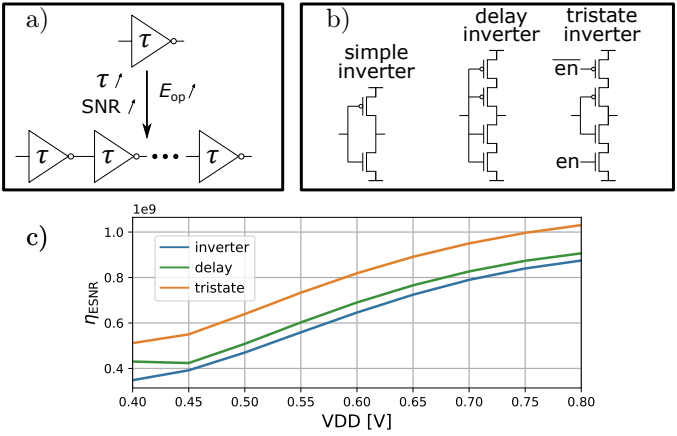

Due to its determinism and the digital encoding, classical static CMOS logic boasts extremely low error rates. 
\Gls{td} computations are inherently prone to noise and variations, thereby introducing errors. 
By choosing a high \gls{snr}, the error can be kept below a certain threshold. 
For selecting a baseline \gls{tdmac} cell, it is therefore important to combine good $\text{SNR}_\text{cell}$ and low energy consumption. 
However, there exists a trade off between the two metrics, making it impossible to find a cell that minimizes both at the same time. 
Cascading $R$ cells increases the SNR of the cascade by a factor of $\sqrt{R}$ while increasing the energy per operation $E_\text{op}$ by a factor of $R$. Therefore, a cell can be improved in SNR at the expense of more energy as indicated in Fig~\ref{fig:cell_comparison}a. 
The comparison of cells can be reduced to the \gls{csnr} as a single metric (Eq. \ref{eq:csnr}), which is  independent of the length $R$ of the cascade. 
\begin{align}
\eta_\text{ESNR}=\frac{\text{SNR}_\text{cell}}{\sqrt{E_\text{op}}}
\label{eq:csnr}
\end{align}
This metric can be used to find a baseline delay element, as a building block for more complex \gls{tdmac} cells. 
Besides using simple inverters, a typical standard cell library offers delay cells, consisting of multiple cascode inverters as displayed in Fig.~\ref{fig:cell_comparison}b (mid). 
As both the input capacitance and the output resistance are doubled, the highest delay per area is achieved here.
The tristate inverter only increases the output resistance, thus consuming less energy than the delay cell inverter while achieving more delay than the simple inverter. 
Circuit simulations of the three cells reveal the benefits of the tristate inverter across a wide applicable voltage range as shown in Fig.~\ref{fig:cell_comparison}c.

Furthermore, voltage scaling can be used in \gls{td} computing to trade off accuracy against energy consumption. However, \gls{tdmac} cells should be designed with nominal voltage in mind, as \gls{csnr} degrades for reduced voltages. 
A good design methodology therefore lies in designing for the highest needed accuracy and reducing $V_\text{dd}$ for less sensitive applications.


In the next step, we look at different \gls{tdmac} cell implementations to determine a baseline cell for the comparison in Section IV.
It can be noted, that designs can be separated to rely on either cascading, slope variation or a combination of the two.
For cascading based implementations, certain parts of the cascade can be skipped, thereby realizing changes in the propagation delay \cite{Everson.2019,Lou.2022,Miyashita.2014}.
As discussed before, the \gls{snr} improves with increasing cascade length, making it applicable to achieve higher accuracy.
For slope variation based implementations, the output slope of an inverter can be varied by either realizing a current starved inverter \cite{Wu.2022,Yang.2021}, increasing output capacitance \cite{Wu.2020,Everson.2021} or by changing the number of parallel transistors in the pull up path \cite{Cai.2021}.
With the exception of increasing output capacitance, slope variation is attractive in terms of energy consumption, as the capacitive load remains constant, leading to no increase in energy.
However, as the mismatch error rises in proportion with the delay step size, the accuracy of slope varying designs has limitations \cite{Chen.2016}.

\begin{figure}[hbt]
\centering
\begin{tikzpicture}
\node at (-4,3.5) {a)};
\node at (-4,-0.5) {b)};
\node at (0,0){\includegraphics[width=0.94\columnwidth]{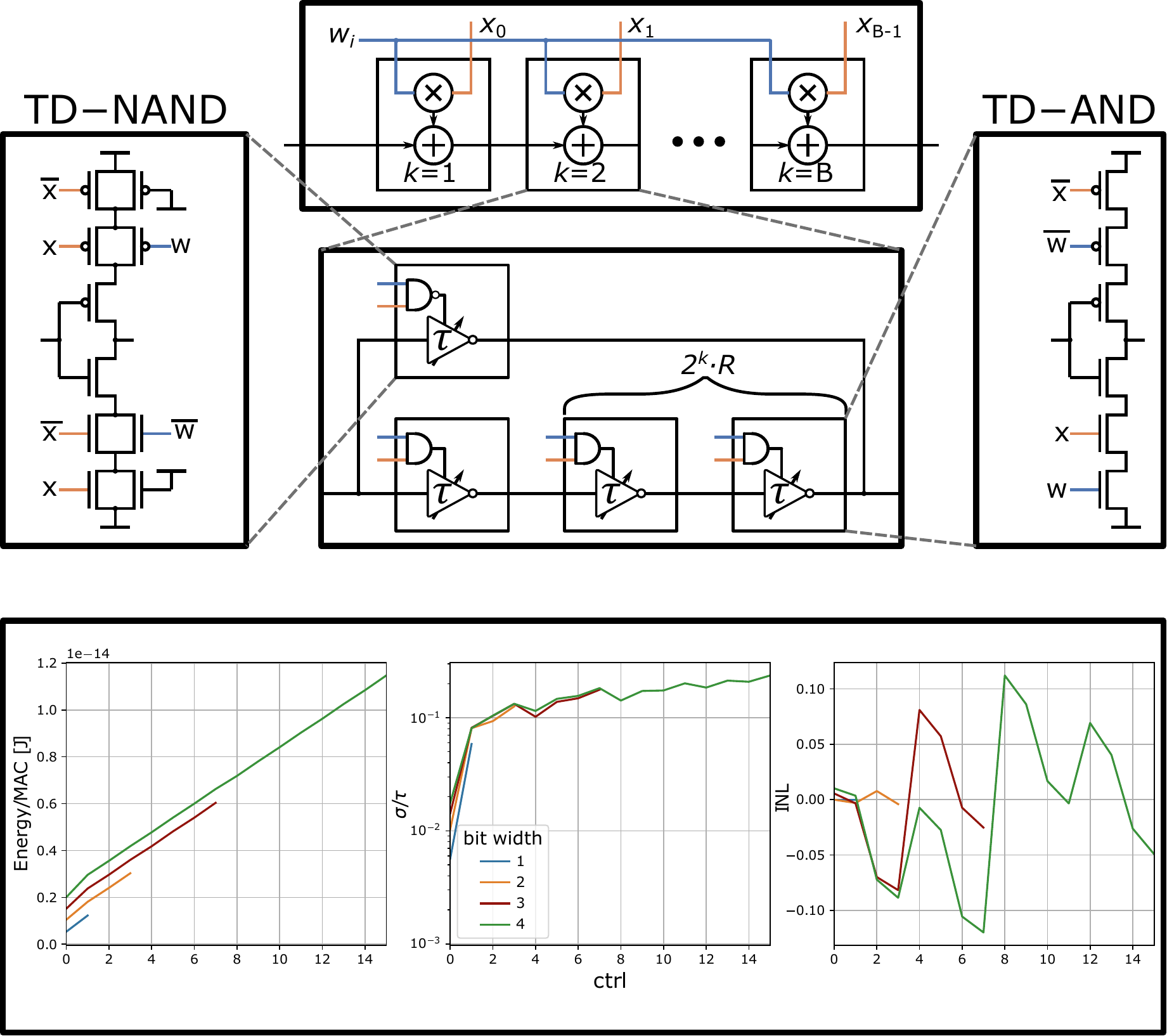}};
\end{tikzpicture}
\caption{Baseline 1xB TDMAC cell. (a) Schematic. (b) Performance metrics.}
\label{fig:mac_cell_final}
\end{figure}

For this investigation, we use the baseline \gls{tdmac} cell given in Fig.~\ref{fig:mac_cell_final}a. 
As we intend to evaluate performance of \gls{td} computing over a large range of array dimensions $N$ and for 1-by-B bit operation, we need a design which is flexible enough to be used over the entire parameter space.
While slope varying cells are feasible for low precision operation, they do not allow high bitlength operation without significantly increasing \gls{inl} and decreasing \gls{snr}.
To keep generality, we therefore use cascading for the \gls{tdmac} cell.
As the tristate inverter has shown the best \gls{csnr} (Fig. \ref{fig:cell_comparison}), we use a tristate like design for the building blocks of the \gls{tdmac} cell which are named {\sc td-and} and {\sc td-nand}. 
The {\sc td-and} cell is only active, if both binary inputs are 1, realizing a logic {\sc and} for the pull up and pull down path. 
The {\sc td-nand} is enabled if any input is 0, thus acting as the bypass path.
Its pull up and pull down resistance is balanced by the added transistor parallel to the always conducting transistor.
While there is a discrepancy between w=0 and w=1, the weight is known a priori for \glspl{ann}, allowing for a calibration of this contribution to nonlinearity.

The performance indicators of the cell can be seen in Fig. \ref{fig:mac_cell_final}b. 
Even for the 4 bit case the \gls{inl} is moderate with the highest peaks only reaching $\pm 0.11$ delay steps.
By increasing the redundancy factor $R$, and therefore increasing the number of cascaded cells per delay step, we can further reduce errors without repeating every component of the circuit.

\section{\bf VMM Array} \label{sec:vmm}
When used in a compute chain, the error of each cell, $\text{err}_\text{cell}(x,w)$, adds to the chain output error, $\text{err}_\text{chain}$.
This leads to a strong relationship between the output error distribution and the input vectors, $X$ and $W$ \cite{freye.2022}.
However, we can use the law of total expectation and the law of total variance, to combine the input dependent cell statistics with the input statistics:
\begin{align}
\mu_\text{err,cell}&=\sum_{i,j}\text{INL}(i,j)\cdot P(x=i) \cdot P(w=j)\\
\sigma_\text{err,cell}^2&=\underbrace{\EX[\text{Var}\left(\text{err}_\text{cell}(x,w)\right)]}_\text{EVPV}+\underbrace{\text{Var}(\text{INL})}_{\text{VHM}=\EX[\text{INL}^2]}\label{eq:expected_val}
\end{align}
Here, the variance of the cell delay can be split into the \gls{evpv}, that consists of the variances for each input combination, and the \gls{vhm} component induced by the INL. Using these cell statistics, we can calculate the statistics for the entire compute chain error:
\begin{align}
    \mu_\text{err,chain}&=N\mu_\text{err,cell}\label{eq:mu_arr}\\
    \sigma_\text{err,chain}^2&=N\cdot \left(\text{EVPV}+\text{VHM}\right)\label{eq:var_arr}
\end{align}

Here, $N$ describes the compute chain length. By increasing the redundancy factor for the delay step $R$ (compare Fig.~\ref{fig:mac_cell_final}), we can reduce all error components.
As the source of nonlinearity stems from the difference between the {\sc td-nand} path and the {\sc td-and} path, which is independent of $R$, $\mu_\text{err,chain}$ is proportional to $\frac{1}{R}$. 
Thus, the \gls{vhm} is proportional to $\frac{1}{R^2}$.
The \gls{evpv} component, on the other hand, shrinks with a factor close to $\frac{1}{R}$, as the mismatch contribution of redundant cells is uncorrelated.
We can summarize the scaling with $R$ in Eq.~\ref{eq:r_scaling}.
\begin{align}
    \mu_\text{err,chain}\propto \frac{1}{R};\quad \text{EVPV}\underset{\sim}{\propto} \frac{1}{R};\quad \text{VHM}\propto \frac{1}{R^2}\label{eq:r_scaling}
\end{align}

To evaluate the performance of a complete \gls{vmm}-array, SPICE simulation results for INL and $\sigma_\text{err,cell}$ are fed into a python framework, which computes Eq. \ref{eq:var_arr} and increases $R$ to make the chain longer and reduce the error below a predetermined threshold.
We assume that the mean error from Eq. \ref{eq:mu_arr} can be calibrated to zero as done in \cite{Lou.2022} and therefore neglect this error contribution.
An additional source of error lies in differences in wire loading. 
Nontheless, this component can be eliminated almost entirely by regular placement and custom routing \cite{Lou.2022}.

The energy consumption for a \gls{mac}, $E_\text{MAC}^\text{TD}$, can be calculated using
\begin{align}
    E_\text{MAC}^\text{TD}=E_\text{cell}+\frac{E_\text{TDC}(N,M)}{N}.
\end{align}
Here, $E_\text{cell}$ is the expected value for the cell energy per operation and includes the increased energy consumption of higher values for $R$. $E_\text{TDC}$ is the energy of the \gls{tdc}.

In the following we assume $\text{err}_\text{chain}$ to be Gaussian shaped and $\text{err}_\text{chain}\leq 3\sigma_\text{err,chain}$. As $x$ and $w$ are given as integer values, no errors are created for $3\sigma_\text{err,chain}\leq~0.5$ due to rounding: $(x\leq0.5)=0$. 
For certain applications that require less accuracy, this threshold can be relaxed to allow lower $R$, increasing efficiency and throughput.

\subsection{\bf Time-to-digital converter}
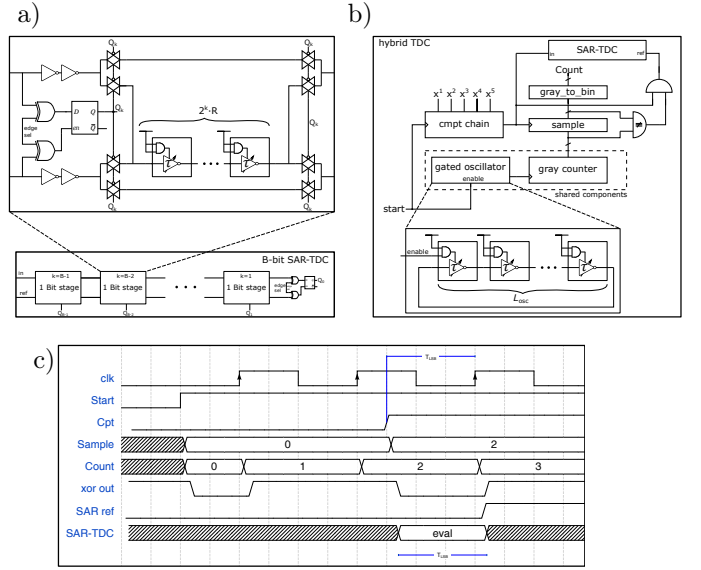
\begin{figure}[htb]
    \centering
    \begin{tikzpicture}
\node at (-4.2,3.8) {a)};
\node at (0.2,3.8) {b)};
\node at (-4.,-0.8) {c)};
    \node at (0,0) {\includegraphics[width=\columnwidth]{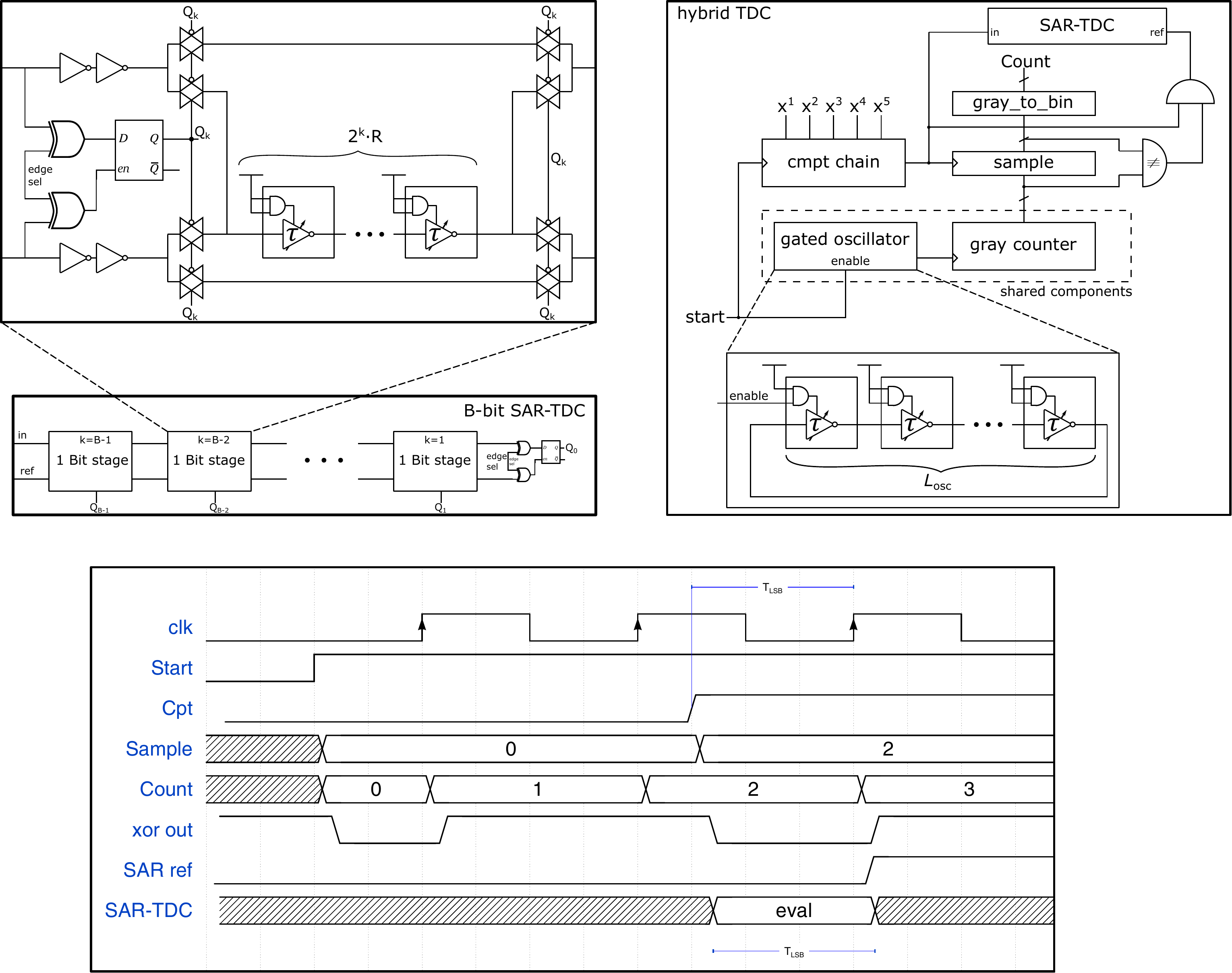}};
    \end{tikzpicture}
    \caption{Schematic for SAR-TDC (a) and hybrid TDC (b). Hybrid TDC working principle (c).}
    \label{fig:tdc_architecture}
\end{figure}

To appropriately estimate the energy consumption for the \gls{tdc}, a corresponding architecture has to be chosen. 
Two common approaches are to build a linear chain of delays that is sampled by a reference signal \cite{Chen.2.2021, Everson.2019, Maharmeh.2021, Song.2021} or to successively delay the faster signal (compute result or reference) in a binary decaying fashion \cite{Szy.2020, Miyashita.2014} (Fig.~\ref{fig:tdc_architecture}a). 
The latter design, also called SAR-TDC, needs less sampling hardware for the same bit width and therefore consumes less energy for larger designs.
To allow for minimum energy and area consumption, the reference input is delayed to arrive at half of the maximum input value, $\frac{\text{max}_\text{in}}{2}$. 
This reduces the binary search to implement a maximum delay of $\frac{\text{max}_\text{in}}{2}$. 
While additional energy has to be spent to delay the threshold signal, it can be shared by all $M$ compute chains running in parallel.
However, the delay within the SAR-TDC still rises exponentially with the bit width, increasing energy consumption and area footprint.

In Fig. \ref{fig:tdc_architecture}b we introduce a novel hybrid \gls{tdc}, which offers better scalability even for large compute chains with high precision.
It comprises a counter, driven by a ring oscillator using $L_\text{osc}$ {\sc td-and} cells to realize a MSB step width, $2L_\text{osc}$, that represents an integer multiple of the unit delay step.
To achieve single unit accuracy, an additional low bit SAR-\gls{tdc} is used for the LSBs, which finds the distance between the compute output and the MSB counter clock. 
An illustration of the working principle is given in Fig. \ref{fig:tdc_architecture}c.
The energy for the counter, $E_\text{cnt}$, and for driving the MSB sampling register, $E_\text{cnt,load}$, add to the total energy, $E_\text{hyb,TDC}$.
However, as the counter and oscillator can be shared among all $M$ compute chains, the hybrid design offers good scaling behavior for large values of $M$ and $N$.
To avoid timing issues when sampling, we use a gray-code counter.
Therefore, only one flipflop input is switched per clock cycle and per compute chain.
The XOR gates in the SAR-\gls{tdc} for the LSB computation allow the \gls{tdc} to operate on rising as well as falling edges.

The counter energy is estimated from synthesis simulations.
We model the energy consumption of the complete \gls{tdc} circuit with Eq. \ref{eq:e_tdc_hyb}.
\begin{equation}
\resizebox{.9\width}{!}{
$\begin{split}
    E_\text{hyb,TDC}&=\left(\frac{E_\text{cnt}}{M}+E_\text{cnt,load}\right)\frac{NR}{2L_\text{osc}}+\frac{2NRE_\text{TD-AND}}{M}\label{eq:e_tdc_hyb}\\
    &+E_\text{TD-AND}2^{\lceil1+\log_2(L_\text{osc})\rceil}+\lceil1+\log_2(L_\text{osc})\rceil E_\text{sample}
    \end{split}$
}
\end{equation}
The optimum oscillator length, $L_\text{osc}$, depends on $N$ and can be determined by finding the minimum of Eq. \ref{eq:e_tdc_hyb}. 
As a simplification, the Gauss brackets are ignored here:

\begin{equation}
\resizebox{.9\width}{!}{
$\begin{split}
    \frac{\delta E_\text{hyb,TDC}}{\delta L_\text{osc}}=2E_\text{TD-AND}+\frac{E_\text{Sample}}{2L_\text{osc}\ln(2)}-\frac{\left(\frac{E_\text{cnt}}{M}+E_\text{cnt,load}\right)NR}{4L_\text{osc}^2}\label{eq:diff_E} \\
    \Rightarrow L_\text{osc}\approx\frac{\sqrt{\left(\frac{E_\text{cnt}}{M}+E_\text{cnt,load}\right)2E_\text{TD-AND}NR\ln(4)}-E_\text{Sample}}{4E_\text{TD-AND}\ln(2)}
    \end{split}$
}
\end{equation}
The energy of the SAR-\gls{tdc} is modeled using Eq.~\ref{eq:e_tdc_sar}:
\begin{equation}
\resizebox{.95\width}{!}{$
    E_\text{SAR-TDC}(B)=E_\text{TD-AND}\frac{M+1}{M}(2^{B}-2)+BE_\text{Sample}\label{eq:e_tdc_sar}
$}
\end{equation}

\begin{figure}[h]
    \centering
    \includegraphics[width=0.9\columnwidth]{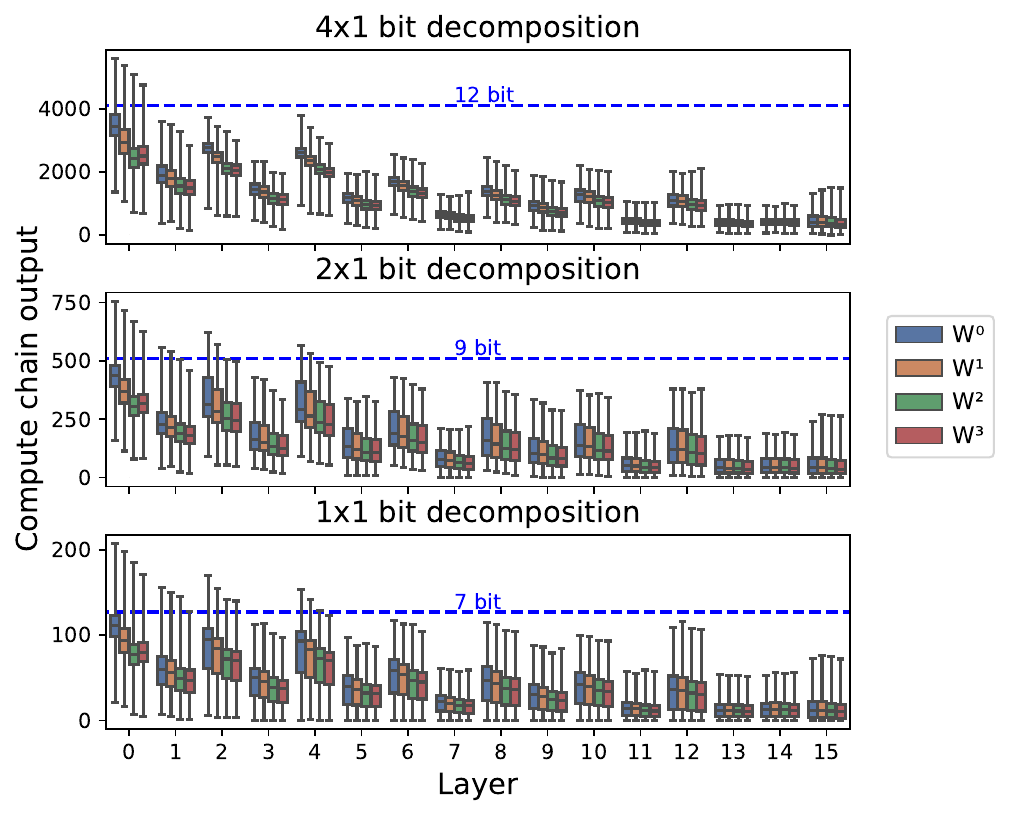}
    \caption{Layer-wise output range of ResNet18 convolution layers for a decomposition into 64 channels.}
    \label{fig:imagenet_range}
\end{figure}
To highlight the scalability of the chosen \gls{tdc} design, we look at the energy consumption for the kernel size of a convolution layer in ResNet18 on the ImageNET dataset in comparison to the SAR approach.
While the channel size increases towards later layers, all channel sizes are integer multiples of 64.
Therefore, we choose 576 (3x3x64) as the baseline compute chain length.
We also look into splitting the channel count even further to 32 and 16.
In \glspl{cnn} we typically observe a limited range of output values. 
By accounting for this behavior, the \gls{tdc} can possibly be reduced in size, which would benefit energy consumption.
In Fig.~\ref{fig:imagenet_range} we look at the layer-wise output range of a ResNet18 inference run on 20 images.
The blue markings are chosen to include most layers full output range with few layers having cut-off outliers.
The channel count was decomposed to 64.
By further reducing the channel count to 32 or 16, the blue line can be reduced further by one or two bits, respectively.
For the number of parallel compute chains, $M$, we assume a value of 8, as used in \cite{Lou.2022}. 
For the decomposed chains, we assume to have a similar silicon area available, hence $M$ can be incresead by a factor of 2 for chain lengths of 288 (3x3x32) and a factor of 4 for chain lengths of 144 (3x3x16).
\begin{figure}[bth]
    \centering
    \includegraphics[width=\columnwidth]{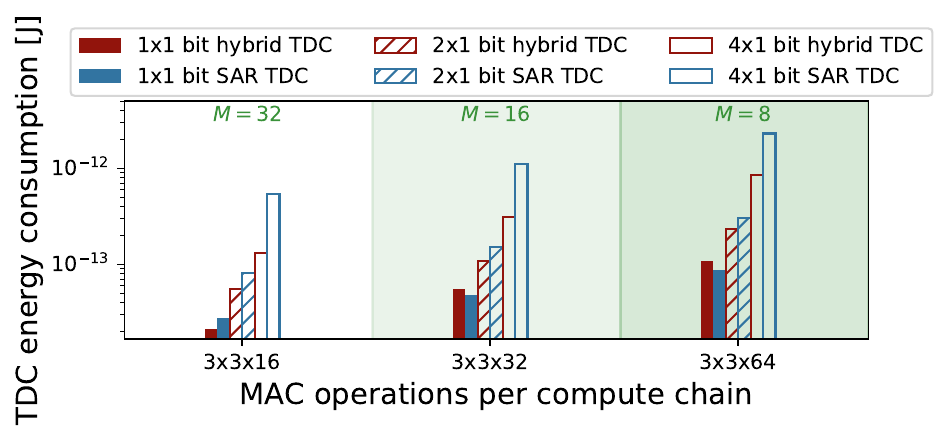}
    \caption{TDC energy for upper limit from Fig. \ref{fig:imagenet_range}.}
    \label{fig:tdc_energy_reduced}
    \vspace{-0.3cm}
\end{figure}

By using the blue markings from Fig.~\ref{fig:imagenet_range} for maximum \gls{tdc} range, the energy consumption was estimated as shown in Fig.~\ref{fig:tdc_energy_reduced}.
For binary operation, the overhead of the counter leads to a lower energy consumption for the SAR approach.
For higher bit widths on the other hand, the scaling behavior of the hybrid approach leads to a clear advantage over the classical approach.
For the upcoming comparison, we will therefore use the hybrid \gls{tdc}.

\section{\bf Comparison} \label{sec:comp}
For a fair comparison, we carefully obtain the energy consumption of the digital and analog implementation. 
For the digital domain, we determine energy consumption for a 1-by-B bit \gls{mac} from post layout simulation results of \gls{vmm} computations at varying precisions and array sizes. 
The \gls{vmm} calculation is performed in a single cycle and is synthesized for \SI{1}{\giga\hertz} operation. 
The accumulation is based on a binary tree of adders and the simulation is performed in TT corner. 
Our approach is to find the energy of the whole array which is then divided through the array length to determine the average energy of the single \gls{mac}.
Similar to the time domain implementation, the weight is fully serialized.

\begin{figure}[htb]
    \centering
    \includegraphics[width=0.8\columnwidth]{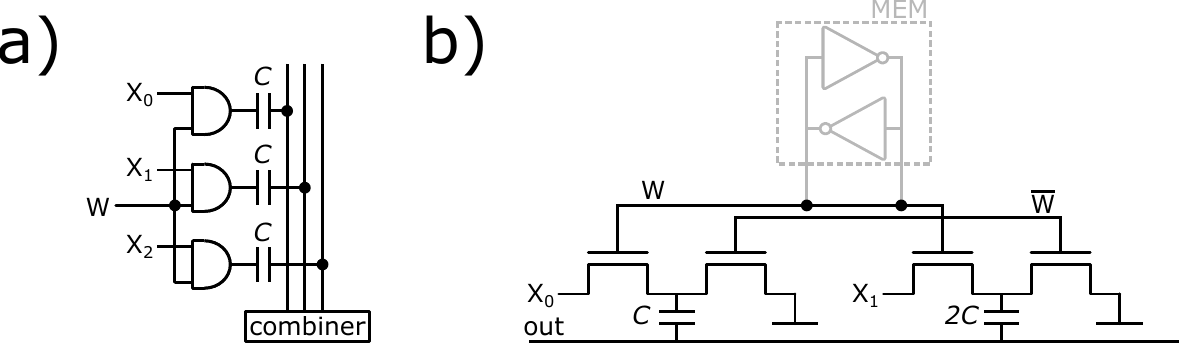}
    \caption{Analog MAC cells. a) Used in \cite{Murmann.2021}. b) Altered from \cite{Sharma.2021}.}
    \label{fig:analog_mac}
\end{figure}

To estimate the energy consumption of the analog domain, a charge domain approach is chosen, as it offers significantly higher \gls{snr} and can therefore be compared more easily to classical digital operation. Energy consumption is estimated similar as given in \cite{Murmann.2021}.
Here, energy per op is estimated using 
\begin{align}
    E_\text{MAC}^\text{ANA}=E_\text{CAP}+E_\text{logic}+\frac{E_\text{ADC}}{N}
\end{align}
with $E_\text{MAC}^\text{ANA}$, $E_\text{ADC}$, $E_\text{CAP}$, $E_\text{logic}$ as the energy of a \gls{mac}, \gls{adc} conversion, accumulation capacitor and the {\sc and}-gate, respectively.
However, by using a similar circuit as presented in \cite{Sharma.2021}, the {\sc and}-gate is reduced to a passthrough transistor, thus eliminating the switching energy of the {\sc and}-gate (Fig.~\ref{fig:analog_mac}).
Another difference to \cite{Murmann.2021}, which is considered here, is the accumulation of charge on only one wire.
This way, the combiner, introducing additional influence of mismatch to the result, can be ommitted and the MSB capacitors are larger, reducing relative mismatch.
Similar to the assumption for the \gls{td}, we introduce a factor $R$, which can be increased, once the mismatch error, $\text{err}_\text{array}$, surpasses a predetermined threshold.

The energy per \gls{adc} conversion, $E_\text{ADC}$, is adopted from \cite{Murmann.2021} and described by Eq. \ref{eq:adc_pow}:
\begin{align}
    E_\text{ADC}=k_1 \text{ENOB}+k_2 4^\text{ENOB}
    \label{eq:adc_pow}
\end{align}
This is obtained by fitting an envelope around a collection of \gls{adc}-designs collected in \cite{adc_survey}. 
Here, we filter out any designs slower than \SI{1}{\mega\hertz} to avoid extremely slow but energy efficient designs to skew the results too much. 
The obtained fitting constants are $k_1$=\SI{0.66}{\pico\joule} and $k_2$=\SI{0.241}{\atto\joule}.

For the comparison, the same 22nm fdSOI technology was used for all three domains.
As it did not offer MIM-capacitors, the use of MOSFET capacitors and simple MOM capacitors were compared.
For the given technology, the relative mismatch in capacitance was much lower for the MOSFET ($<$2.5\%). 
The non-linearity of the capacitance is assumed to be compensatable, thus the energy consumption and mismatch of the MOSFET were used for evaluating the analog domain. 

\begin{figure}[bth]
    \centering
    \includegraphics[width=\columnwidth]{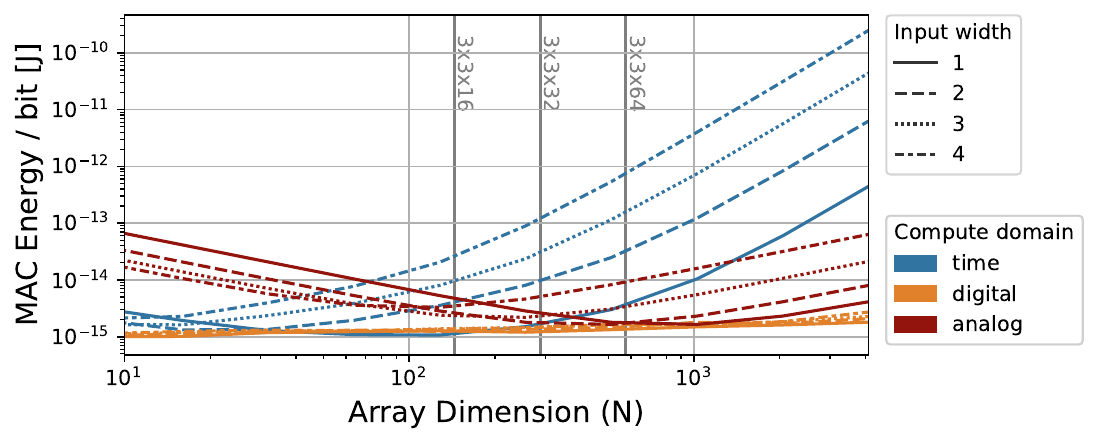}
    \caption{Energy comparison of different array dimensions for all three domains for $\text{err}_\text{chain}\leq 0.5$.}
    \label{fig:energy_comp}
\end{figure}

For the weights, the bitwise sparsity of a ResNet18 inference was investigated to lie between 60\% and 80\%. 
To correctly estimate energy consumption, a sparsity of 70\% is therefore considered for the weights.
The result of the energy comparison can be seen in Fig.~\ref{fig:energy_comp}. 
Aside from few exceptions, the digital domain implementation dominates both \gls{td} as well as analog domain throughout the complete range.
The main reason for this is the accuracy requirement, $\text{err}_\text{chain}\leq 0.5$.
For the \gls{td} implementation, this mainly leads to an increase of $R$, raising the energy consumption but also lowering the throughput.
For higher input widths, the \gls{tdmac} cell cascades more delay cells and thus shows larger errors. 
Factor $R$ therefore has to be set higher to stay below the set error limit. This leads to higher energy consumption for higher input widths.
In the analog domain, the tight accuracy restrictions not only increase $R$, but also increase the required \gls{adc} \gls{snr}.

\begin{figure}[htb]
    \centering
    \begin{tikzpicture}
    \node[anchor = south] at (0,0) (p1){
    \includegraphics[width=\columnwidth]{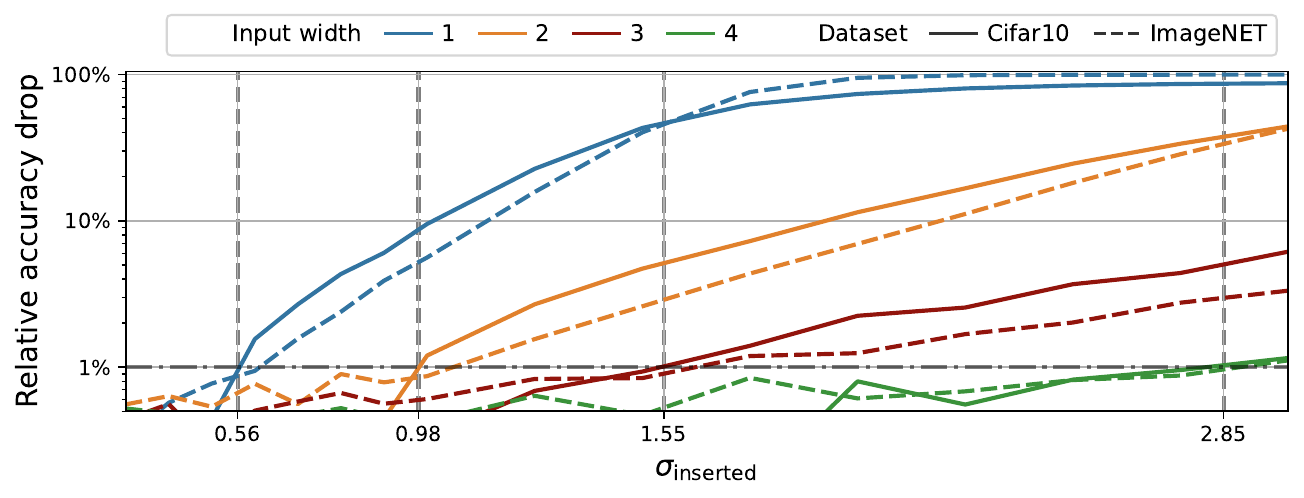}};
    \node[anchor = north] at (0,0) (p2){\begin{tabular}{c|c|c|c|c}
         MAC cell & 1-by-1 & 1-by-2 & 1-by-3 & 1-by-4\\
         \hline
         $\sigma_\text{array,max}$ & 0.58 & 0.98 & 1.55 & 2.85
    \end{tabular}};
    \node[anchor = east] at (-4,3.4){a)};
    \node[anchor = east] at (-4,0){b)};
    \end{tikzpicture}
    \caption{a) Relative accuracy drop ($1-\frac{\text{Acc}(\sigma=\sigma_\text{inserted})}{\text{Acc}(\sigma=0)}$) over inserted noise (lower is better). b) Selected maximum noise.}
    \label{fig:network_sensitivity}
\end{figure}
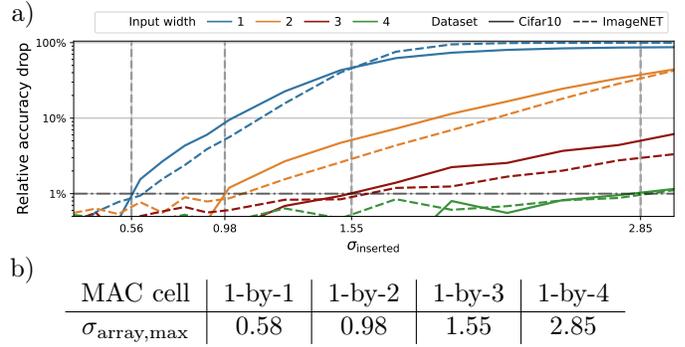

As the maximum error is set below the quantization limit, the same accuracy is reached for all three domains. 
However, the regions where the analog as well as the \gls{td} shine the most are error-resilient applications. 
In Fig. \ref{fig:network_sensitivity}a, we introduce noise to inference runs of ResNet20 on the Cifar10 dataset and ResNet18 on the ImageNET dataset, which are both quantized to 4 bit using \gls{lsq}\cite{esser.2020}.
The noise is Gaussian shaped and introduced to the convolution result according to the necessary bit sequencing for the corresponding \gls{tdmac} cell.
Rounding is then applied to account for \gls{tdc} conversion.
We define acceptable noise levels, $\sigma_\text{array,max}$, by limiting the tolerated relative accuracy drop to $\leq1\%$. 
The crossing point with this threshold gives us $\sigma_\text{array,max}$ (Fig.~\ref{fig:network_sensitivity}b). This crossing occurs earlier for Cifar10 classification.

Using this information, we can relax the requirements for the application example. 
By allowing a standard deviation corresponding to Fig.~\ref{fig:network_sensitivity}b, we can not only reduce $R$, but also reduce the requirements for the \gls{adc} by calculating the needed ENOB given by Eq. \ref{eq:enob}.
\begin{align}
    \text{ENOB}=\frac{\text{SNR}-1.76}{6.02}
    \label{eq:enob}
\end{align}

Under these conditions, \gls{td} and analog computing both become more competitive in terms of energy consumption and surpass the digital implementation.
While the \gls{td} design dominates for small array sizes, increasing length also increases noise and therefore $R$, which drives energy consumption in the \gls{td}.
The charge domain on the other hand benefits from the cost of the \gls{adc} increasing slower than the amount of \glspl{mac}, thus energy consumption decreases with array dimension.

Considering the precision, analog computing benefits from higher wordlengths due to more efficient sharing of the \gls{adc} up to high array dimensions. As \gls{td} computing is less dominated from the \gls{tdc} energy, the benefit of partially sharing the \gls{tdc} is outweighed by the increase in energy consumption and error susceptibility of designs with higher input width.
\begin{figure}[htb]
    \centering
    \includegraphics[width=\columnwidth]{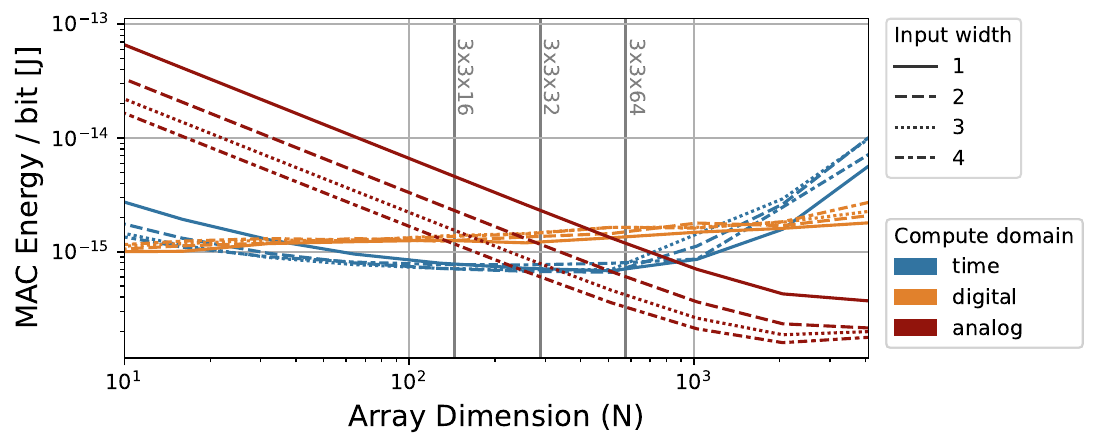}
    \caption{Energy comparison of different array dimensions for all three domains for $\sigma_\text{arrray}$ according to Fig.~\ref{fig:network_sensitivity}b.}
    \label{fig:energy_comp_2}
    \vspace{-0.25cm}
\end{figure}

\subsection{\bf Throughput and Area}
\begin{figure}[htb]
    \centering
    \begin{tikzpicture}
    \node[anchor = south] at (0,0) (p1){\includegraphics[width=\columnwidth]{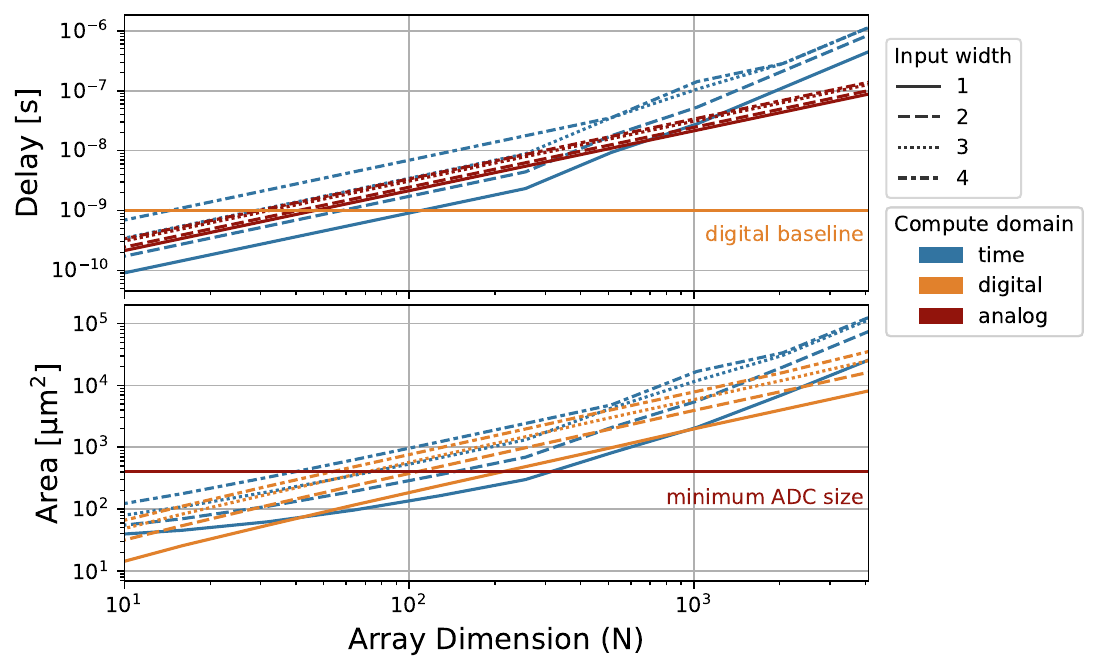}};
    \node[anchor = south] at (-4.2,5.1){a)};
    \node[anchor = south] at (-4.2,2.6){b)};
    \end{tikzpicture}
    \caption{Comparison of different array dimensions for all three domains for $\sigma_\text{arrray}$ according to Fig.~\ref{fig:network_sensitivity}b: a) Throughput. b) Area.}
    \label{fig:troughput_area}
\end{figure}

Considering throughput, which is displayed in Fig.~\ref{fig:troughput_area}a, the digital domain seems to be the most attractive option for medium to high array sizes.
While the analog and \gls{td} implementations seem to outperform for smaller array sizes, by employing pipelining or using lower threshold devices, the digital implementation can be tuned to even faster operation speeds.
For the analog implementation we assume a shared \gls{adc} with $M=8$.
The resulting throughput is lower than the \gls{td} design for low wordlengths and small to medium array sizes. 
For higher array sizes and greater wordlengths, the \gls{adc} scales better and starts to dominate the \gls{td} implementation.
Similar to the energy estimation, the \gls{adc} throughput was determined by building an envelope of the data from \cite{adc_survey}. Here, the filter for the data was extended to exclude designs with too high energy consumption (three times the energy in Eq. \ref{eq:adc_pow}).

The area of the \gls{tdmac} cell presented in Fig.~\ref{fig:mac_cell_final} can be easily estimated as two subcells of type TD-AND or TD-NAND can always be combined in one Euler path, thus one diffusion break per two subcells has to be added to the number of transistors. The cell area can therefore be estimated by
\begin{align}
    A_\text{cell}=\left(B\cdot9+7\cdot R\cdot\sum_{i=0}^B2^i\right)\cdot\text{CPP}\cdot H_\text{cell},
\end{align}
with CPP representing the contacted poly pitch and $H_\text{cell}$ giving the standard cell height.
The area for the \gls{tdc} comprises the area of the used TD-AND cells, the area of other standard cells and the synthesis result of the grey-code counter.
For the digital domain evaluation, the area after \gls{pnr} is obtained.

Analog computing scales poorly with the technology node.
Besides SRAM cells shrinking slower than standard cells, the major reason for this is the \gls{adc}, which uses long channel devices to ensure proper performance.
As one \gls{adc} is shared with multiple compute arrays, the area footprint is reduced at the cost of throughput.
To get a fair comparison, the same filter as for the throughput was applied with \glspl{adc} with insufficient \gls{snr} to compute arrays larger than 100 \glspl{mac} also being filtered out.
From this selection, the smallest design was chosen.

The results of the comparison are shown in Fig.~\ref{fig:troughput_area}b.
We can observe, that the classical digital approach offers best area use for smaller array sizes, while the \gls{td} implementation is on a par from certain array sizes for lower input word widths.
The \gls{adc} area consumption scales better than the two alternatives, allowing for smaller overall design for high arrray dimensions.
\section{\bf Conclusion}\label{sec:conc}
In this work, we quantitavely compare classical digital computing with computing in the analog and the time domain.
The evaluation is first performed for error free computation and then adjusted to errors, which can be tolerated by modern \glspl{ann}.
We introduce a novel \gls{tdmac} cell and a novel \gls{tdc} as a generic and scalable baseline which achieves high accuracy over a large range of array sizes and bit lengths.
For this, we present a new efficiency metric which combines energy efficiency and mismatch accuracy.

For error free computation, the digital design proves dominant, as reducing error probability requires too high SNR from the analog and \gls{td} implementation. 
Considering error tolerance, the necessary bit wise accuracy for ResNet20 as well as ResNet18 was investigated.
By backannotating this accuracy, the energy consumption of the analog as well as time domain implementation can be drastically reduced.
While the digital domain still outperforms considering throughput, the \gls{td} implementation is more efficient for small to medium sized vector sizes and the analog domain implementation dominates large vector sizes.
In terms of area requirements, \gls{td} generally is not competitive.
To conclude, it can be said that \gls{td} computing remains promising for all small to medium size \gls{vmm} applications with certain error tolerance which are not too constrained in area. As the industry keeps scaling down classical CMOS technology nodes, \gls{td} computing will keep improving in terms of energy consumption and required area. 
High throughput applications on the other hand remain dominated by classical digital implementations.

\bibliographystyle{IEEEtran}
\bibliography{IEEEabrv,ref}

\end{document}